\documentclass[english,keywords,aps,amsmath,amssymb,twocolumn]{revtex4-1}
\usepackage[T1]{fontenc}
\usepackage[latin1]{inputenc}
\usepackage{babel}
\usepackage{amssymb}
\usepackage{graphicx}
\usepackage{color}
\usepackage{bm}
\usepackage{longtable}
\usepackage{amsmath} 
\usepackage{amsfonts}
\usepackage{amssymb}
\begin{document}
\title{L$\ddot{u}$der rule, von Neumann rule and Cirelson's bound of Bell CHSH inequality }
\author{Asmita Kumari$^{1,2}$ \footnote{
asmitakumari@hri.res.in}}
\author{A. K. Pan $^{2}$\footnote{akp@nitp.ac.in}}
\affiliation{$^{1}$Harish-Chandra Research Institute, HBNI, Chhatnag Road, Jhunsi, Allahabad 211 019, India}
\affiliation{$^{2}$National Institute of Technology Patna, Ashok Rajhpath, Patna 800005, India}
\begin{abstract}
In [PRL, 113, 050401 (2014)] the authors have shown that instead of L$\ddot{u}$der rule, if degeneracy breaking von Neumann projection rule is adopted for state reduction, the quantum value of three-time Leggett-Garg inequality can exceed it's L$\ddot{u}$ders bound. Such violation of  L$\ddot{u}$ders bound may even approach algebraic maximum of the inequality in the asymptotic limit of system size. They also claim that for Clauser-Horne-Shimony-Holt (CHSH) inequality such violation of L$\ddot{u}$ders bound (known as Cirelson's bound) cannot be obtained even when the measurement is performed sequentially first by Alice followed by Bob. In this paper, we have shown that  if von Neumann projection rule is used, quantum bound of CHSH inequality exceeds it's Cirelson's bound and may also reach its algebraic maximum four. This thus provide a strong objection regarding the viability of von Neumann rule as a valid state reduction rule. Further, we pointed out that the violation of Cirelson's bound occurs due to the injection of additional quantum non-locality by the act of implementing von Neumann measurement rule.

\end{abstract}

\maketitle

\section{Introduction}
Bell theorem \cite{bell64} provides one of the most fundamental results in the field of quantum foundations. Through the quantum violation of a suitable set of inequalities, Bell demonstrated that a theory which respects the notion of local realism is in contradiction with quantum mechanics. The simplest Bell's inequality is Clauser-Horne-Shimony-Holt (CHSH) form defined for two-party, two-measurement and two-outcome scenario \cite{chsh69}. The maximum quantum value of CHSH expression is $2 \sqrt{2}$, known as  Cirelson's bound \cite{cri}. This bound is achieved for maximally entangled states and local anti-commuting observables. However, there are no-signaling post-quantum theories that violate Cirelson's bound, but such predictions have not been seen in nature \cite{popescu}.

Of late, the study of macrorealism and Leggett-Garg inequalities \cite{lg85,halliwell16,swati17,pan17,pan20} has received considerable attention. It is known that for dichotomic observables, using L$\ddot{u}$ders projection rule \cite{lu}, the maximum quantum bound of a three-time Leggett-Garg inequalities \cite{lg85,halliwell16,swati17,pan17} is $1.5$, irrespective of the system size \cite{budroni13}. But, Budroni and Emary \cite{budroni14} have shown that instead of L$\ddot{u}$der rule, if degeneracy breaking von Neumann projection rule is used for the measurement of dichotomic observables in a $d$-dimensional ($d>2$) system, the quantum violation of Leggett-Garg inequalities can exceed the L$\ddot{u}$ders bound (temporal Cirelson's bound) and can approaches algebraic maximum in the asymptotic limit of system size.

In this regard, the natural question is whether similar violation of Cirelson's bound of CHSH inequality can also be achieved using von Neumann state update rule if the measurements are performed sequentially first by Alice and followed by Bob. Budroni and Emary \cite{budroni14} claimed that it is not possible due to commutativity of measurements involved in correlations in CHSH inequality. The post measurement state remains same for both the case of using L$\ddot{u}$ders and von Neumann state update rule and hence  Cirelson's bound is the maximum achievable bound.  Note that, we have recently questioned the implication of von Neumann projection rule for the cases of the violation of  L$\ddot{u}$ders bound of Leggett-Garg inequalities and non-contextual inequalities \cite{AQA}. We argued that, the quantum violation of Leggett-Garg inequalities by invoking the von Neumann rule should not be treated as the traditional notion for the quantum violation Leggett-Garg inequalities.

 In contrast to the assertion in \cite{budroni14}, in this paper, we demonstrate that the use of von Neumann rule for state reduction indeed provides quantum value of CHSH expression exceeding the Cirelson's bound. Such a result is clearly not acceptable. We thus provide a stronger objection against von Neumann rule to be considered as a viable projection rule. In order to demonstrate our result, we studied the quantum violation of CHSH inequality for entangled state as well as for product state. We found that if von Neumann rule is used for state reduction then for a maximally entangled state and local anti-commuting observables, the quantum value of CHSH expression exceeds the Cirelson's bound. Like the case of Leggett-Garg inequalities mentioned earlier, the violation of Cirelson's bound of CHSH inequality may also reach its algebraic maximum for particular choice of basis. Moreover, we show that even when Alice and Bob share product state, the violation of CHSH inequality can be obtained using von Neumann state update rule. Since both the result are not acceptable then something goes terribly wrong here at all. We provide the reason of such violations and what goes wrong in implementing von Neumann rule.  We argued that a strong form of 'quantum non-locality' is introduced by the act of implementing von Neumann rule by Alice which is not possible for physically separated systems.

This paper is organized as follows. In Sec.II, we show the difference between L$\ddot{u}$ders and von Neumann rules in sequential measurement of two degenerate observables of $(\mathbb{C}^2 \otimes \mathbb{C}^2)$ system. In Sec.III,  we explicitly demonstrated the violation of Cirelson's bound of CHSH inequality using von Neumann projection rule for maximally entangled state and local anti-commuting observables.  Further, we showed that use of von Neumann rule leads to the violation of local bound of CHSH inequality even for product state. We discuss the reason of such violation of Cirelson's bound in Sec. IV.

\section{ Difference between L$\ddot{u}$ders and von Neumann rules in Bell's correlation scenario}
Let us first encapsulate the essence of the L$\ddot{u}$ders rule and the von Neumann rule proposed in \cite{budroni14}.
For a degenerate observable there are two ways to implement the state reduction; in one case state is reduced to degenerate subspace ( L$\ddot{u}$ders rule) and in another scenario to each eigenstate (von Neumann rule). 

Consider an observable $\hat{O}$ having discrete eigenvalues $o_1, o_2, o_3$... $o_m$ with degree of degeneracies $x_1,x_2,x_3$...$x_m$ respectively. Let  $P_{m}^{\alpha}=|\phi_{m}^{\alpha}\rangle\langle\phi_{m}^{\alpha}|$ is the projection operator associated with $m^{th}$ eigenvalue  where $\alpha$ denotes the degeneracy. The von Neumann projection rule breaks the degeneracy, so that, the reduced density matrix can be written as
\begin{eqnarray}
\rho_{v}=\sum_{m,\alpha}{P_{m}^{\alpha}\rho P_{m}^{\alpha}}
\end{eqnarray}
where   $\rho$ is the initial density matrix of the system. As already indicated, $\rho_{v}$ is \textit{not} unique for degenerate observable. On the other hand, the L$\ddot{u}$ders projection rule respects the degeneracy. The reduced density matrix in this case can be written as 
\begin{eqnarray}
\rho_{l}=\sum_{m}{P_{m}\rho P_{m}}
\end{eqnarray}
where $P_{m}=\sum_{\alpha=1}^{x_{m}}|\phi_{m}^{\alpha}\rangle\langle\phi_{m}^{\alpha}|$  \cite{heger, pan}. Then, for an observable with degenerate eigenvalues, the von Neumann rule provides the reduced density matrix less coherent than that is obtained using the L$\ddot{u}$ders rule. For non-degenerate observable both the rules are identical. Throughout our paper by von Neumann rule we refer the discussion in this paragraph to avoid any confusion.

Let  $\hat{A} \otimes \mathbb{I}$ and $\mathbb{I} \otimes \hat{B}$ be two dichotomic degenerate observables in $\mathbb{C}^2 \otimes \mathbb{C}^2$, such that, $A^{\alpha}_{m}$ and $B^{\beta}_{n}$ are their respective projectors. Here $m,n=\pm 1$ are the eigenvalues and $\alpha, \beta = 1, 2$ are degeneracy of  $\hat{A} \otimes \mathbb{I}$ and $\mathbb{I} \otimes \hat{B}$  respectively. The joint expectation value of  $\hat{A} \otimes \mathbb{I}$ and $\mathbb{I} \otimes \hat{B}$  can be written as
\begin{eqnarray}
\nonumber
{\langle \hat{A} \otimes \hat{B}\rangle} = {\langle (\hat{A} \otimes \mathbb{I}) (\mathbb{I} \otimes \hat{B}) \rangle}_{seq}= \sum_{m,n=\pm {1}}mn P(m, n)
\\
\label{Eq.p1}
\end{eqnarray}
where the tensor product ${\langle \hat{A} \otimes \hat{B}\rangle}$ can also be seen as sequential measurement of $(\hat{A} \otimes \mathbb{I})$ and $(\mathbb{I} \otimes \hat{B})$. If the joint probability $P(m,n)$ is calculated by using L$\ddot{u}$ders rule, then
\begin{equation}
P^{l}(m,n) = Tr[{{{A_m}}}\rho {A_m}{B_n}]
\end{equation}  
where ${A_m}=\sum_{\alpha}A^{\alpha}_m$, $B_n=\sum_{\beta}{B^{\beta}_n}$ and $\rho$ is the state shared by Alice and Bob. This means that while implementing the state reduction via L$\ddot{u}$ders rule there is no need to specify degeneracies $\alpha$ and $\beta$. However, for our degenerate observables $\hat{A} \otimes \mathbb{I}$ and $\mathbb{I} \otimes \hat{B}$, the joint probability using von Neumann projection rule is obtained as
\begin{equation}
P^{v}(m,n)= Tr \bigg[\sum_{\alpha,\beta}{A^{\alpha}_{{m}}} \rho A^{\alpha}_{m} B^{\beta}_{n}\bigg]
\end{equation}
For subsystem in $\mathbb{C}^2 \otimes \mathbb{C}^2$, let us define the observable $\hat{A} \otimes \mathbb{I}= A^{1}_{+} + A^{2}_{+}- A^{1}_{-}- A^{2}_{-}$ and $\mathbb{I} \otimes \hat{B}= B^{1}_{+} + B^{2}_{+}- B^{1}_{-}- B^{2}_{-}$. Here $A^{1}_{+}$ and $A^{2}_{+}$ are the degenerate projectors with $m=+1$ and  $A^{1}_{-}$ and $ A^{2}_{-}$ are the degenerate projectors with $m=-1$. Similarly for the projectors of $\mathbb{I} \otimes \hat{B}$. The joint probability of $A_{+} = A^{{1}}_{+} +  A^{{2}}_{+}$ and $B_{+} = B^{1}_{+} + B^{2}_{+}$ using L$\ddot{u}$ders rule can be written as
\begin{equation}
P^{l}(+ +) = Tr[(A^{{1}}_{+} +  A^{{2}}_{+}) \rho (A^{{1}}_{+} + A^{{2}}_{+})(B^{1}_{+} + B^{2}_{+})]
\end{equation}
Using joint probabilities for other combinations of eigenvalues given in Eq.({\ref{Eq.p1}}), the joint expectation value is obtained by L$\ddot{u}$ders rule is ${\langle \hat{A} \otimes \hat{B}\rangle}^{l} = Tr[\rho \hat{A} \otimes \hat{B}]$.

Using von Neumann rule the joint probability $P(+ +)$ can be written as
\begin{equation}
\label{v}
P^{v}(+ +) = Tr[(A^{{1}}_{+}\rho A^{{1}}_{+} + A^{{2}}_{+}\rho A^{{2}}_{+})(B^{1}_{+} + B^{2}_{+})]
\end{equation}
On the other hand, using joint probabilities for other combinations of eigenvalues given in Eq.({\ref{Eq.p1}}),  the joint expectation value using von Neumann  rule is obtained as
\begin{eqnarray}
\label{lv}
{\langle \hat{A} \otimes \hat{B}\rangle}^{v} = {\langle \hat{A} \otimes \hat{B}\rangle}^{l} &-& Tr[(A^{1}_{+}\rho A^{2}_{+} + A^{2}_{+}\rho A^{1}_{+} \\ \nonumber &-& A^{1}_{-}\rho A^{2}_{-}- A^{2}_{-}\rho A^{1}_{-})\mathbb{I} \otimes \hat{B}] \  \  \
\end{eqnarray} 
Comparing Eq.(\ref{lv}) with Eq.(\ref{v}) it is seen that there exists an additional term in Eq.(\ref{lv}) along with ${\langle \hat{A} \otimes \hat{B}\rangle}^{l}$. Note that, the additional term, $Tr[(A^{1}_{+}\rho A^{2}_{+} + A^{2}_{+}\rho A^{1}_{+} - A^{1}_{-}\rho A^{2}_{-}- A^{2}_{-}\rho A^{1}_{-})(\mathbb{I} \otimes\hat{B})]$ is in general nonzero for degenerate observables and may depend on the choice of basis. However, ${\langle \hat{A} \otimes \hat{B}\rangle}^{l}$ always remain basis independent.

\section{Apparent violation of Cirelson's bound of CHSH inequality}
Consider the Bell scenario in which two specially separated observers Alice and Bob are allowed to do local measurements on the shared state in their respective site. If $A_1$ and $A_2$ are the observables belonging to Alice and, $B_1$ and $B_2$ are for Bob, then the CHSH inequality is given by
\begin{eqnarray}
\label{cbell}
\Delta = \langle A_{1}B_{1}\rangle+\langle A_{1}B_{2}\rangle+\langle A_{2}B_{1}\rangle-\langle A_{2}B_{2}\rangle\leq2
\end{eqnarray}
Here $\langle A_{1}B_{1}\rangle$ stands for $\langle A_{1} \otimes B_{1}\rangle$ and similarly for others. Note that, the correlation $\langle A_{1}\otimes B_{1}\rangle$ can also be calculated sequentially, so that, $\langle A_{1} \otimes B_{1}\rangle = \langle (A_{1}\otimes \mathbb{I}) (\mathbb{I} \otimes B_{1})\rangle_{seq}$. This means that Alice first measures a degenerate observable $A_{1}\otimes \mathbb{I}$ and then Bob perform the measurement by $\mathbb{I} \otimes B_{1}$. Note that, $[A_{1}\otimes \mathbb{I}, \mathbb{I} \otimes B_{1}] = 0$. Both the observables, $A_{1}\otimes \mathbb{I}$ and $\mathbb{I} \otimes B_{1}$ are degenerate observable having $\pm 1$ eigenvalues with two eigenstates corresponding to each eigenvalue. 

We consider the following choices of local anticommuting observables which provide Cirelson's bound of CHSH expression, so that, $A_1=  \sigma_{x} \otimes \mathbb{I} $, $A_2=  \sigma_{z} \otimes \mathbb{I} $, $ B_1=  \mathbb{I}\otimes \frac{\sigma_{z}-\sigma_{x}}{\sqrt{2}}$ and $B_2=\mathbb{I}\otimes \frac{\sigma_{z}+\sigma_{x}}{\sqrt{2}} $. Decomposing the observable $A_1$ in terms of its projectors, we obtain
\begin{eqnarray} 
\label{a1l}
\nonumber
A_1=  \sigma_{x} \otimes \mathbb{I}  &=& A^{1}_{1+} + A^{2}_{1+}- A^{1}_{1-}- A^{2}_{1-} \\ 
&=& |a^{1}_{1+}\rangle \langle a^{1}_{1+}| + |a^{2}_{1+}\rangle \langle a^{2}_{1+}| \\ \nonumber
 &-& |a^{1}_{1-}\rangle \langle a^{1}_{1-}| - |a^{2}_{1-}\rangle \langle a^{2}_{1-}|
\end{eqnarray}
where, $A^{1}_{1+}=|a^{1}_{1+}\rangle \langle a^{1}_{1+}|$ and  $A^{2}_{1+}=|a^{2}_{1+}\rangle \langle a^{2}_{1+}|$ are the projectors of $A_1$ corresponding to $+1$ eigenvalue, and $A^{2}_{1-} = |a^{1}_{1-}\rangle \langle a^{1}_{1-}|$ and $ A^{2}_{1-}= |a^{2}_{1-}\rangle \langle a^{2}_{1-}|$ are the projectors corresponding to $-1$ eigenvalue. Here, $ |a^{1}_{1+}\rangle  = (0,1,0,1)^{T}/\sqrt{2}$, $|a^{2}_{1+}\rangle = (1,0,1,0)^{T}/\sqrt{2}$, $|a^{1}_{1-}\rangle = (0,-1,0,1)^{T}/\sqrt{2}$ and $|a^{2}_{1-}\rangle = (-1,0,1,0)^{T}/\sqrt{2}$ are the eigenvectors of $A_1$ having eigenvalues $(1,1,-1,-1)$ respectively.
 Since choice of basis for implementing von Neumann rule is not unique, then Alice is free to choose any suitable basis of the same observable $A_1$ in calculating a various joint correlations with Bob. Considering a general basis, observable $A_1$ can also be decomposed as
\begin{eqnarray} 
\label{vona1}
\nonumber
A_1&=& A^{'1}_{1+} + A^{'2}_{1+}- A^{'1}_{1-}- A^{'2}_{1-} \\ 
&=& |a^{'1}_{1+}\rangle \langle a^{'1}_{1+}| + |a^{'2}_{1+}\rangle \langle a^{'2}_{1+}| \\ \nonumber
 &-& |a^{'1}_{1-}\rangle \langle a^{'1}_{1-}| - |a^{'2}_{1-}\rangle \langle a^{'2}_{1-}|
\end{eqnarray}
where $|a^{'1}_{1+}\rangle = \eta_1 |a^{1}_{1+}\rangle + \sqrt{1-\eta^{2}_1}|a^{2}_{1+}\rangle$,  $|a^{'2}_{1+}\rangle = \sqrt{1-\eta^{2}_1} |a^{1}_{1+}\rangle - \eta_1 |a^{2}_{1+}\rangle$,  $|a^{'1}_{1-}\rangle = \gamma_1 |a^{1}_{1-}\rangle + \sqrt{1-\gamma^{2}_1}|a^{2}_{1-}\rangle $ and  $|a^{'2}_{1-}\rangle = \sqrt{1-\gamma^{2}_1}|a^{1}_{1-}\rangle - \gamma_1|a^{2}_{1-}\rangle $ with $\eta_1, \gamma_1 \in [0, 1]$.\\

Note that, $A_{1+} = A^{'1}_{1+} + A^{'2}_{1+} = A^{1}_{1+} + A^{2}_{1+}$.
Similarly, the observable $A_2$ can also be decomposed as 
\begin{eqnarray} 
\label{a2l}
\nonumber
A_2=  \sigma_{z} \otimes \mathbb{I}  &=& A^{1}_{2+} + A^{2}_{2+}- A^{1}_{2-}- A^{2}_{2-} \\ 
&=& |a^{1}_{2+}\rangle \langle a^{1}_{2+}| + |a^{2}_{2+}\rangle \langle a^{2}_{2+}| \\ \nonumber
 &-& |a^{1}_{2-}\rangle \langle a^{1}_{2-}| - |a^{2}_{2-}\rangle \langle a^{2}_{2-}|
\end{eqnarray}
where $ |a^{1}_{2+}\rangle  = (0,1,0,0)^{T} $, $|a^{2}_{2+}\rangle =(1,0,0,0)^{T} $, $|a^{1}_{2-}\rangle = (0,0,0,1)^{T}$ and $|a^{2}_{2-}\rangle = (0,0,1,0)^{T}$ are the eigenvectors of $A_2$ with eigenvalues $(1,1,-1,-1)$ respectively. Again, decomposing $A_2$ using general basis one has
\begin{eqnarray} 
\label{vona2}
\nonumber
A_2&=& A^{'1}_{2+} + A^{'2}_{2+}- A^{'1}_{2-}- A^{'2}_{2-} \\ 
&=& |a^{'1}_{2+}\rangle \langle a^{'1}_{2+}| + |a^{'2}_{2+}\rangle \langle a^{'2}_{2+}| \\ \nonumber
 &-& |a^{'1}_{2-}\rangle \langle a^{'1}_{2-}| - |a^{'2}_{2-}\rangle \langle a^{'2}_{2-}|
\end{eqnarray}
where $|a^{'1}_{2+}\rangle = \eta_2 |a^{1}_{2+}\rangle + \sqrt{1-\eta^{2}_2}|a^{2}_{2+}\rangle$,  $|a^{'2}_{2+}\rangle = \sqrt{1-\eta^{2}_2} |a^{1}_{2+}\rangle - \eta_2 |a^{2}_{2+}\rangle$,  $|a^{'1}_{2-}\rangle = \gamma_2 |a^{1}_{2-}\rangle + \sqrt{1-\gamma^{2}_2}|a^{2}_{2-}\rangle $ and  $|a^{'2}_{2-}\rangle = \sqrt{1-\gamma^{2}_2}|a^{1}_{2-}\rangle - \gamma_2|a^{2}_{2-}\rangle $  with $\eta_2, \gamma_2 \in [0, 1]$.\\
 Using the above mentioned decompositions of $A_1$ and $A_2$ given by Eqs.(\ref{vona1}) and Eqs.(\ref{vona2}), the detailed calculation of CHSH expression for entangled and product states are presented.

 \subsection{ Quantum value of CHSH expression for entangled State}
Let Alice and Bob share an entangled state given by
\begin{eqnarray}
| \Psi \rangle_{E} = \frac{1}{\sqrt{2}}(\sin{\alpha}, -\sin{\beta}, \cos{\beta}, cos{\alpha})^{T}
\end{eqnarray}
which becomes maximally entangled state for $\alpha = \beta = \pi/4$. For aforementioned choices of observables, using L$\ddot{u}$ders rule maximum quantum value of CHSH inequality is obtained to be $2 \sqrt{2}$, which is the Cirelson's bound. We are now interested in the joint correlations calculated by using von Neumann rule instead of L$\ddot{u}$ders rule. Note here that Alice may choose different basis to perform the measurement of a particular observable (say, $A_1$) by considering any choices of basis to implement state reduction through von Neumann rule. Here, for the measurement of $A_1 $, Alice uses the basis by taking the parameters $\eta_1$ and $ \gamma_1$ in Eq.(\ref{vona1}) to calculate correlation $\langle A_{1}B_{1}\rangle$. But for calculating $\langle A_{1}B_{2}\rangle$, she takes the parameters $\eta'_1$ and $ \gamma'_1$. In general, $\eta_1 \neq \eta'_1$ and $\gamma_1 \neq \gamma'_1$. Similarly when Alice measures observable $A_2$, for the correlation $\langle A_{2}B_{1}\rangle$ she takes $\eta_2$ and $ \gamma_2$ in Eq.(\ref{vona2}). But for calculating $\langle A_{2}B_{2}\rangle$ she chooses different parameter $\eta'_2$ and $ \gamma'_2$ with $\eta_2 \neq \eta'_2$ and $\gamma_2 \neq \gamma'_2$ in general. The, quantum value of CHSH expression using von Neumann rule is obtained as
 \begin{widetext}
 \begin{eqnarray}
 \nonumber
  {(\Delta^v_E)}_Q& =&\frac{1}{\sqrt{2}}\Bigg[2+\gamma_1  \left(2 \gamma_1 \left(\gamma_1 \left(\sqrt{1-\gamma_1^2}+\gamma_1\right)-1\right)-\sqrt{1-\gamma_1^2}\right)-\gamma_2 \left(\sqrt{1-\gamma_2^2}+2 \gamma_2 \left(\gamma_2^2-\gamma_2\sqrt{1-\gamma_2^2} -1\right)\right)\\ \nonumber
   &+&\eta'_1 \left(\sqrt{1-{\eta_1}^{'2}}+2 \eta'_1 \left({\eta_1}^{'2}-\eta'_1\sqrt{1-{\eta_1}^{'2}} -1\right)\right)+{\eta_2}^{'2} \left(\sqrt{1-{\eta_2}^{'2}}-2 \eta'_2 \left(\eta'_2 \left(\sqrt{1-{\eta_2}^{'2}}+\eta'_2\right)-1\right)\right)\\ \nonumber
    &+&\eta_1 \left(2 \eta_1 \left(\eta_1 \left(\sqrt{1-\eta_1^2}+\eta_1\right)-1\right)-\sqrt{1-\eta_1^2}\right)-\eta_2 \left(\sqrt{1-\eta_2^2}+2 \eta_2 \left(\eta_2^2-\sqrt{1-\eta_2^2} \eta_2-1\right)\right)\\ 
     &+&\gamma'_1  \left(2 \gamma'_1 \left(\gamma'_1 \left(-\sqrt{1-{\gamma_1}^{'2}}+\gamma'_1\right)-1\right)+\sqrt{1-{\gamma_1}^{'2}}\right)+\gamma'_2 \left(\sqrt{1-{\gamma_2}^{'2}}-2 \gamma'_2 \left({\gamma_2}^{'2}+\gamma'_2\sqrt{1-{\gamma_2}^{'2}} -1\right)\right)\Bigg]
 \end{eqnarray}
 \end{widetext}
We plotted the quantum CHSH expression ${(\Delta^v_E)}_Q$ against $\gamma_1$ (a specific choice of basis) in Figure.1, by taking $\gamma'_1 = 0.20$, $\eta_1 = 0.98$, $\eta'_1 = 0.20$,  $\eta_2 = 0.83$, $\eta'_2 = 0.57$, $ \gamma_2 = 0.83$ and $\gamma'_2 = 0.57$.
\begin{figure}[ht]
\begin{minipage}[c]{0.5\textwidth}
\includegraphics[width=1\linewidth]{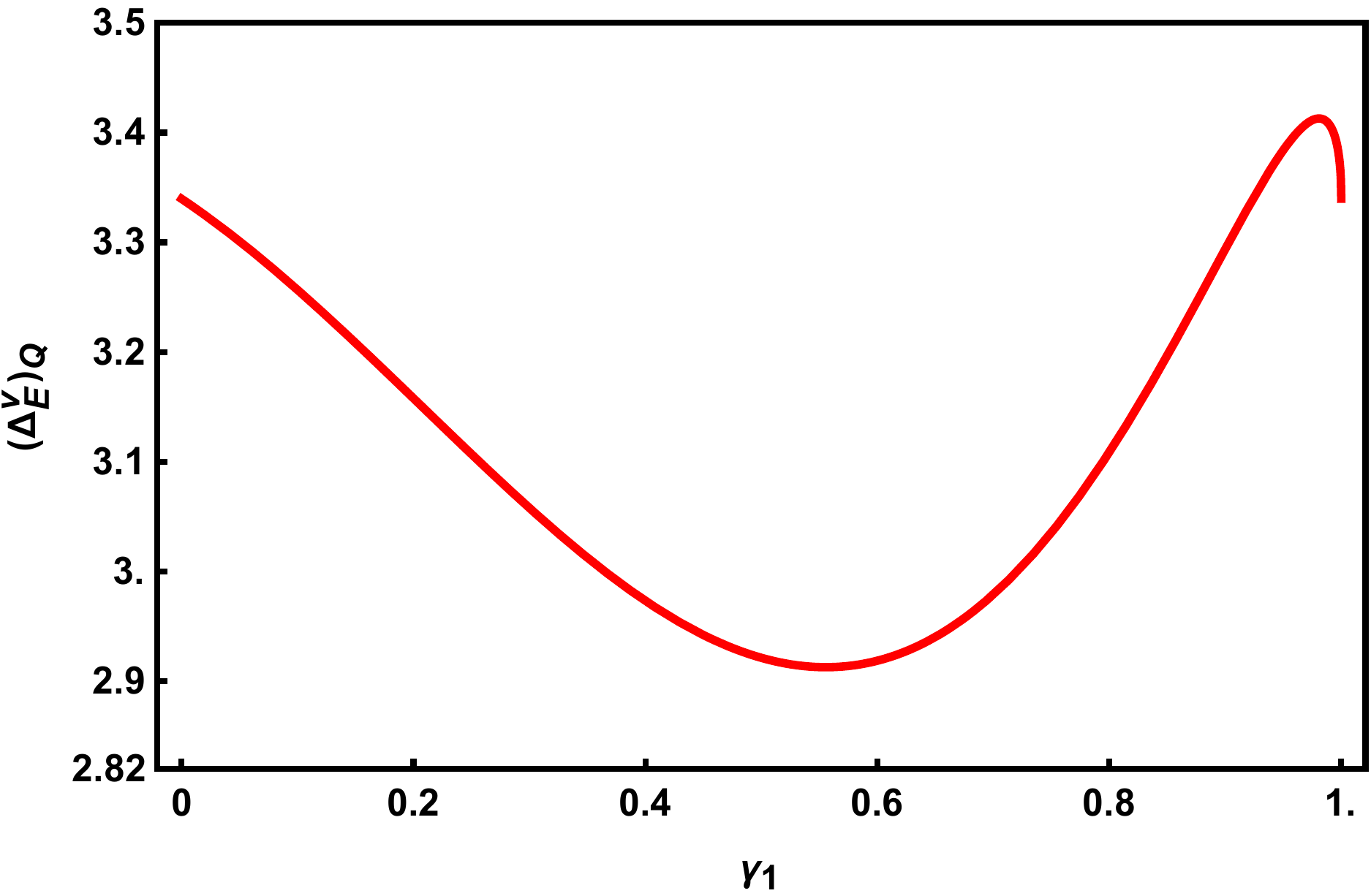}
\end{minipage}\hfill
\begin{minipage}[hc]{0.5\textwidth}
\caption{(color online): The quantity ${(\Delta^v_E)}_Q$ is plotted against $\gamma_1$ (see text for details).}
\label{fig:01}
\end{minipage}
\end{figure}
The quantum value of CHSH expression for the above choices of parameters is found to be ${(\Delta^v_E)}_Q \approx 3.41$, which is larger than the Cirelson's bound ($2\sqrt{2} \approx 2.82$). Thus the von Neumann rule provides an improved violation of CHSH inequality beyond the Cirelson's bound. This is in contrast to the claim by Budroni and Emary \cite{budroni14}, who argued that the violation of Cirelson's bound is not possible through the von Neumann rule even for the measurement is performed sequentially. We shall come back to the discussion regarding why such an wierd violation of CHSH inequality is found.

 We shall now show even more counter-intuitive results, i.e., the violation of CHSH inequality for product state by using von Neumann rule.

\subsection{ Quantum value of CHSH expression for product state}
Let us assume that the state shared between Alice and Bob is product state given by
\begin{eqnarray}
| \Phi \rangle_{P} = N_{2}(\sin{\alpha}, \sin{\alpha}, \cos{\beta}, cos{\beta})^{T}
\end{eqnarray}
where $N_{2} = (2 (\sin^2(\alpha) + \cos^2(\beta)) )^{-1/2}$ is the normalization constant. For the system state $|\Phi \rangle_{P}$ and aforementioned choices of observable using L$\ddot{u}$ders rule, the quantum expression of CHSH inequality is obtained as 
\begin{eqnarray}
{(\Delta^l_P)}_Q = \frac{\cos (2 \alpha )+\cos (2 \beta )}{\sqrt{2} \left(\sin ^2(\alpha )+\cos ^2(\beta )\right)} 
\end{eqnarray}
 The maximum quantum value of ${(\Delta^l_P)}_Q $ is obtained to be $\sqrt{2}$ for $\alpha =0$ and $\beta = \pi/4$. Since we are interested in the correlations calculated by von Neumann rule, following the similar procedure in the case of entangled state, the quantum expression of CHSH inequality for system state $|\Phi \rangle_{P}$ with $\alpha =0$, $\beta = \pi/4$, and aforementioned choices of observables is given by
\begin{widetext}
\begin{eqnarray}
\nonumber
 {(\Delta^{v}_P)}_Q& =& \frac{1}{\sqrt{2}}\Bigg[2-\gamma_1 \left(\sqrt{1-\gamma_1^2}+2 \gamma_1 \left(\gamma_1^2-\gamma_1\sqrt{1-\gamma_1^2} -1\right)\right)+\left(2 \gamma_2^2-1\right) \left(2 \gamma_2 \left(\sqrt{1-\gamma_2^2}-\gamma_2\right)+1\right)\\ \nonumber 
 &+&\eta'_1 \left(\sqrt{1-\eta_1^{'2}}-2 \eta'_1 \left(\eta'_1 \left(\sqrt{1-\eta_1^{'2}}+\eta'_1\right)-1\right)\right)+\eta_1 \left(\sqrt{1-\eta_1^2}+2 \eta_1 \left(\eta_1^2-\eta_1 \sqrt{1-\eta_1^2} -1\right)\right)\\  
  &+&\gamma'_1 \left(2 \gamma'_1 \left(\gamma'_1 \left(\sqrt{1-\gamma_1^{'2}}+\gamma'_1\right)-1\right)-\sqrt{1-\gamma_1^{'2}}\right)-\left(2 \gamma_2^{'2}-1\right) \left(2 \gamma'_2 \left(\sqrt{1-\gamma_2^{'2}}+\gamma'_2\right)-1\right)\Bigg]N
\end{eqnarray}
\end{widetext}
The CHSH expression $ {(\Delta^{v}_P)}_Q$ for the product state $|\Phi \rangle_{P}$ at  $\alpha =0$, $\beta = \pi/4$ is plotted against basis $\gamma_1$ in Figure.2, by taking $\eta_1 = 0.20$,  $\eta'_1 = 0.55$,$\gamma'_1 = 0$, $\eta_2 = 0.85$, $\eta'_2 = 0$, $ \gamma_2 = 0.85$ and $\gamma'_2 = 0.55$.
\begin{figure}[ht]
\begin{minipage}[c]{0.5\textwidth}
\includegraphics[width=1\linewidth]{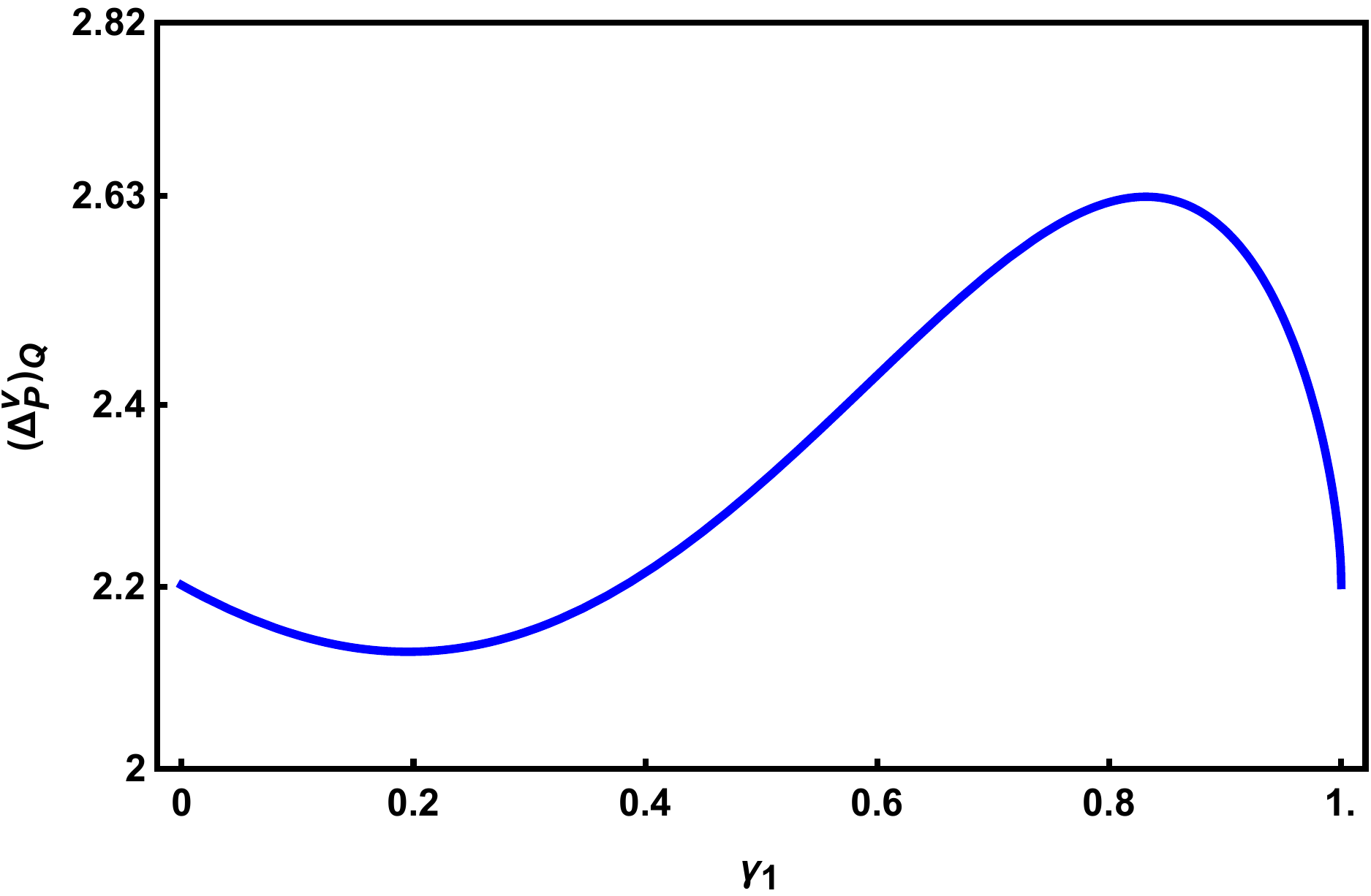}
\end{minipage}\hfill
\begin{minipage}[hc]{0.5\textwidth}
\caption{(color online): The quantity ${(\Delta^{v'}_P)}_Q$ is plotted against $\gamma_1$ at $\alpha =0$, $\beta = \pi/4$ (see text for details).}
\label{fig:02}
\end{minipage}
\end{figure}

 From Figure. 2, it is found that if von Neumann rule is adopted, the quantum value of CHSH expression $ {(\Delta^{v}_P)}_Q$ is $2.63$ which is greater than local bound $2$. This result is truly not acceptable as it is widely known that the quantum value of CHSH expression beyond $2$ demonstrates the signature of non-locality and entangled state is necessary. Thus, from the above two examples one can simply conclude that there is something terribly wrong in the von Neumann rule and needs to be rejected.

\section{Discussion}
For a degenerate observable there are apparently two ways to implement the state update rule. In one case the state is reduced to degenerate subspace ( L$\ddot{u}$ders rule) corresponding to the eigenvalues and in other case to each of the eigenstates (von Neumann rule). In this paper, we provided a strong objection against the von Neumann rule to be considered as a viable state reduction rule. In contrast to the claim made by Budroni and Emary \cite{budroni14} we have shown that inspite of commutativity of measurements involved in CHSH inequality if performed sequentially, the von Neumann state update rule can provide the quantum value of CHSH expression ($\Delta_{Q}$) beyond Cirelson's bound for entangled states as well as product states. 

We now explain how the implementation of von Neumann rule introduces an additional  non-locality to the quantum correlations. In particular, we demonstrate that how Alice's measurements influence the Bob's measurements due to the act implementing the von neumann rule. Let $\rho_{AB}$ is the initial state of the system shared by Alice and Bob. Consider the case when Alice performs the measurement of the observable $A_1$ and uses the von Neumann rule for state reduction. From Eq.(\ref{vona1}) we can say that the state will be reduced to each of the eigenstates of $A_1$.  The observable $A_1 = \sigma_x \otimes \mathbb{I}$ can be written as $A_1 = A_{1 +} \otimes \mathbb{I}-A_{1 -} \otimes \mathbb{I}$, where $A_{1 \pm} = (\mathbb{I} \pm \sigma_x)/2$. If Alice implement L$\ddot{u}$ders rule, she used $A_{1 +} \otimes \mathbb{I}$ corresponding to the eigenvalues $\pm 1$ for state reduction. But, if Alice wants to implement the von Neumann rule by considering general basis defined in Eq.(\ref{vona1}) she has to reduce the state by considering the projectors $A^{'1}_{1+}$ and $A^{'2}_{1+}$ where $A_{1+}\otimes \mathbb{I} =A^{'1}_{1+}+A^{'2}_{1+}$. Interestingly, one can write 
 \begin{eqnarray}
 \nonumber
A^{'1}_{1+}= A_{1+}\otimes C_{+}; \ \ ;  A^{'2}_{1+}= A_{1+}\otimes (\mathbb{I} - C_{+})
 \end{eqnarray}
where $C_{+} = \eta_1 \sqrt{1-\eta^2_1}\sigma_x + (1-\eta^2_1)|0\rangle \langle 0|+\eta^2_1|1\rangle \langle 1|$. Then, $C_{+}$ is actually acting non-locally on the particle which is in possession to Bob and is physically not possible. But, notion of von Neumann rule forced us to implement this kind of measurements. In other words the act of implementing von Neumann rule introduces an additional quantum non-local effect from Alice to Bob. This, in fact, enables the violation of Cirelson's bound of CHSH inequality, which has no relevance to the usual violation of CHSH inequality by quantum theory.  Thus, the von Neumann rule is not a viable projection rule and should be rejected.

\section*{Acknowledgments}
 AKP acknowledge the support from the project DST/ICPS/QuEST/Theme 1/2019/4.

\end{document}